**Development of a DNA sensor using molecular logic gate**

D. Bhattacharjee, Dibyendu Dey, S. Chakraborty, Syed Arshad Hussain*

Thin Film and Nanoscience Laboratory

Department of Physics, Tripura University, Suryamanianagar – 799022, Tripura, India

S. Sinha

Department of Botany, Tripura University, Suryamanianagar – 799022, Tripura, India

* Corresponding author
Email: sa_h153@hotmail.com, sah.phy@tripurauniv.in
Phone: +91 381 2375317 (O)
Fax: +91 381 2374802 (O)

**Abstract** This communication reports the increase in fluorescence resonance energy transfer (FRET) efficiency between two laser dyes in presence of Deoxyribonucleic acid (DNA). Two types of molecular logic gates have been designed where DNA acts as input signal and fluorescence intensity of different bands are taken as output signal. Use of these logic gates as DNA sensor has been demonstrated.

# 1 Introduction

Recent research on molecular logic gate was initialized by the recognition of the pioneering work of de Silva [1] using the beguiling idea that molecules can manipulate and process information using logic, as do electronic computers and human brains. A remarkable progress has been made since then in the development of molecular logic gates [7-19]. The change in fluorescence characteristics of a dye due to the introduction of some external agent can be considered to be analogous to the digital responses in electronic logic gates. Molecules can undergo changes in ground or exited state due to the interference of some external chemical or biological materials [2]. In most of the cases, these kinds of changes can be realized in terms of the basic operations of logic gates using the familiar Boolean logic [3]. The increasing role of electronic devices in our daily life, as well as our constant need to pursue superior technologies, have raised a wide interest among the researchers in the development of molecular systems mimicking the operation of electronic logic gates and circuits not only as simple logic gates but also as more complex devices [4-6] such as adders/subtractors [7-9], multiplexers/demultiplexers [10-12], encoders/decoders [13-14], keypad locks [15-19] etc. Among many applications of molecular logic gates, one of the most interesting one is the investigation of the inside components of a cell, where silicon-based analogues are not expected to reach [7]. To have some structural and functional ideas about different biological materials like DNA, RNA, proteins etc the development of techniques for sensing and monitoring them is in great demand [20]. A homogeneous sensing ensemble, based on DNA quadruplexes was reported by D.Margulies et al [21] where the sensing ensemble can generate as well as provide a direct analysis of the properties of the target proteins.

In designing nano-sized devices the issue of connectivity plays an important role. The strength of semiconductor devices depends on this connectivity where the output from one nano-sized device is used to control the input of another. Whereas in case of molecular logic gate it is not easy to pass the output of one gate to serve as an input to the next due to the difference in nature and properties of output input characteristics. From this point of view it must be stressed that molecular logic gate or computation need not follow the conventional semiconductor blue print.

Using some basic logic, the molecules can process and manipulate information as like electronic computers and human brain. There are many molecular logic gates where chemicals are used as inputs and optical signals are the outputs [22]. For the sensing of different organic [23], inorganic [24] and biological [21, 25] materials these molecular logic gates are now being extensively used. For the easier understanding of the outputs of the sensors they are compared with some well known digital logic gates and from different outputs of those logic gates we can have some idea about the different structural features of the test sample.

DNA is an interesting anionic polyelectrolyte with unique double helix structure whose base sequence controls the heredity of life [26]. K. Fujimoto et al reported the detection of target DNAs by excimer-monomer switching of Pyrene using the Fluorescence resonance energy transfer (FRET) process [27]. DNA based nanomachine was reported by H. Liu et al using the FRET phenomenon [20]. Also for encrypting messages on DNA strands, various methods have been accomplished [28-30].

Present communication reports the effect of DNA on fluorescence resonance energy transfer (FRET) between two dyes Acriflavine (Acf) and Rhodamine B (RhB). Acf and RhB are in principle suitable for energy transfer. Both the dyes are highly fluorescent. Fluorescence spectrum of Acf overlaps with absorption spectrum of RhB. By using this FRET process we are able to construct a photo-regulated fluorescence switch. The output of the switch is mimicking the electronic NOT and YES/NOT logic gates. This kind of "ON-OFF" switching of fluorescence intensity can be varied by the introduction of photochemically active biomolecule DNA. It has been observed that the incorporation of DNA in the FRET pair modulates the FRET efficiency. This has been used to design the molecular logic gate, which is capable of sensing the presence of DNA.

**2 Experimental Section**

Both Acriflavine (Acf) and Rhodamine B (RhB) were purchased from Sigma Chemical Co., USA and were used as received. Ultrapure Milli-Q water (resistivity 18.2 MΩ-cm) was used as solvent. The DNA used is sheared Salmon sperm DNA having a size of nearly about 2000 bp with approximate GC content 41.2%., purchased from SRL India and was used as received. The purity of DNA was checked by UV-Vis absorption and fluorescence spectroscopy before use. UV–Vis absorption and fluorescence spectra of the solutions were recorded by a Perkin Elmer

Lambda-25 absorption spectrophotometer and Perkin Elmer LS-55 Fluorescence Spectrophotometer, respectively. For fluorescence measurement the excitation wavelength was 420 nm. The concentration of the individual dye in aqueous solution was $10^{-6}$ M. In order to have the mixed dye solution the dye solutions were mixed with 1:1 volume ratio.

**3 Results and discussions**

3.1 FRET between Acf and RhB in aqueous solution

Figure 1 shows the fluorescence spectra of pure Acf (curve 1), pure RhB (curve 2) and their mixture of 1:1 volume ratio (curve 3) in aqueous solution. Spectra shown in figure 1 were recorded with excitation wavelength 420 nm (close to the monomer absorption of Acf). This excitation wavelength was selected in order to avoid the direct excitation of the RhB molecules. With this excitation wavelength Acf shows prominent fluorescence with peak at 500 nm (curve-1 of figure 1), whereas, RhB fluorescence intensity is almost negligible with a very weak peak at 578 nm (curve-2 of figure 1). From the spectral characteristics it has been observed that both the Acf and RhB are mainly present as monomer in aqueous solution. However for the fluorescence spectra of Acf-RhB mixed solution (curve 3), the RhB fluorescence intensity increases even with this excitation wavelength (420 nm) as well as Acf fluorescence decreases compared to their pure counterpart. This may be due to transfer of energy from Acf to RhB. This transferred energy excites more RhB molecules followed by light emission from RhB, which is added to the original RhB fluorescence. As a result the RhB fluorescence intensity gets sensitized. In order to confirm this we measure the excitation spectra with emission wavelength fixed at Acf (500 nm) and RhB (578 nm) fluorescence maximum in case of Acf-RhB mixed aqueous solution (inset 2 of figure 1). Interestingly both the excitation spectra are almost similar and possess characteristic absorption bands of Acf monomers. This confirms that the RhB fluorescence in case of Acf – RhB mixed solution is mainly due to the light absorption by Acf and corresponding transfer to RhB monomer. Thus FRET between Acf to RhB has been confirmed.

3.2 FRET between Acf and RhB in presence of DNA

Figure 2 shows the fluorescence spectra of Acf - RhB mixed aqueous solution (1:1 volume ratio) in presence (curve 2) and in absence (curve 1) of DNA. The DNA concentration was 1 μg/ml. It is interesting to observe that in presence of DNA the RhB fluorescence intensity increases and the Acf fluorescence intensity decreases further compared to that in absence of DNA. This indicates that presence of DNA influence the extent of energy transfer.

Based on the fluorescence spectra of figure 1 and figure 2 the fluorescence energy transfer efficiency have been calculated using the following equation [31]

$$E = 1 - \frac{F_{DA}}{F_D}$$

Where $F_{DA}$ is the relative fluorescence intensity of the donor in the presence of acceptor and $F_D$ is the fluorescence intensity of the donor in the absence of the acceptor.

It has been observed that the FRET efficiency of the dye pair increases from 11.37% (absence of DNA) to 79.1% (presence of DNA). These data support the increase in energy transfer between Acf and RhB in presence of DNA. It is interesting to mention in this context that FRET process is distance dependent and if the inter molecular distance between donor and acceptor decreases, then the transfer of energy from donor to acceptor becomes much efficient. FRET is effective over distance ranging in between 1 and 10 nm [32]. Also increase in spectral overlap integral enhances the energy transfer [33-34].

DNA contains two long polymer strands and repeating units called nucleotides or bases [35-36]. The bases lie horizontally between the two spiraling polymer strands with negatively charged phosphate backbones attached on either side of the base pair [35-36]. The distance between two consecutive base pairs is 0.34 nm [37]. In present case both the dyes Acf and RhB used are cationic. In presence of DNA they are attached with the DNA strands through the electrostatic attraction with the negatively charged phosphate backbone of DNA. As a result the both the dyes come close to each other resulting favorable condition for energy transfer. Accordingly, the energy transfer efficiency increases in presence of DNA. Attachment of the dyes onto the phosphate backbone of DNA has been shown schematically in figure 3. It may be mentioned in this context that Shu Wang et al reported that the negatively charged DNA bring a close electrostatic interaction with the cationic water soluble conjugated polymer backbone referring to an efficient FRET [38]. DNA strands have also been used in FRET-based biosensors, where they are used as spacers between FRET dye pairs.

3.3 Design of molecular logic gates

Based on the efficiency of FRET between Acf and RhB in presence and absence of DNA, two types of molecular logic gates have been proposed, namely NOT and YES/NOT gates. These molecular logic gates, unlike digital counter parts; sense the presence of a biological material DNA which acts as an input signal. The output signal is the fluorescence intensity of a particular band (500 nm and 578 nm). Using these logic gates it is possible to detect the DNA in aqueous solution up to a very low concentration of 1 µg/ml.

3.4 Design of NOT gate as DNA sensor

Based on the spectral characteristic in figure 2 we have designed the logic gates. Here we consider the fluorescence intensity of 500 nm band during FRET between Acf and RhB, as the output signal and presence of DNA as input. Fluorescence intensity of 400 units (shown in figure-2) has been chosen as the reference level. Table 1 shows the logic of NOT gate. In the absence of DNA (input = 0), fluorescence intensity at 500 nm band is greater than the reference level (output = 1). In presence of DNA (input = 1) the 500 nm fluorescence band intensity is less than the reference level (output = 0). Thus an effective NOT gate can be developed which can sense the presence of DNA in aqueous solution having concentration as low as 1 µg/ml. Thus by observing the fluorescence intensity of 500 nm band it is possible to detect the presence of DNA.

3.5 Design of YES-NOT gate as DNA sensor

In this case the input is similar to that of NOT gate, where as the output signals are the fluorescence intensities of 500 nm and 578 nm bands. When the input signal is zero (absence of DNA) the intensity of 500 nm band is greater than the reference level (output = 1) and for 578 nm band the fluorescence intensity is less than the reference level (output = 0). When input signal is one (presence of DNA) the output of 500 nm band is 0 where as 578 nm band is 1. In this case the YES-NOT gate confirms strongly the presence and absence of DNA in the aqueous solution. Table 2 shows the logic of YES-NOT gate. Here by comparing the intensity of 500 nm with the reference level it is possible to detect the presence or absence of DNA.

It is worthwhile to mention in this context that in the present manuscript the experiments have been done with sheared Salmon sperm DNA having a size of nearly about 2000 bp with approximate GC content 41.2%. The actual size of the genomic DNA is approximately $3\times 10^9$ bp which is sheared to 2000 bp. Therefore the sheared DNA in solution contains huge number of different kinds of sequences. In order to check the dependence of experimental results on the specific sequences of DNA we have also tested the whole experiments with isolated human DNA (GC content 40%) and found similar results (result not shown). Therefore, the working principle of the designed logic gate depends on the interaction of Acf-RhB with phosphate moiety of DNA and is independent of any specific sequences of DNA. Using this designed logic gates only the presence or absence of DNA can be detected.

**4 Conclusions**

Based on the experimental observation that the presence of DNA increase the fluorescence resonance energy transfer (FRET) between two laser dyes Acriflavine (Acf) and Rhodamine B (RhB) two types of molecular logic gates, namely, NOT and YES-NOT gate have been designed. These two molecular logic gates have been found efficient to detect the presence of DNA in aqueous solution having concentration as low as 1 µg/ml.

**Acknowledgements**

The author SAH is grateful to DST, CSIR and DAE for financial support to carry out this research work through DST Fast-Track project Ref. No. SE/FTP/PS-54/2007, CSIR project Ref. 03(1146)/09/EMR-II and DAE Young Scientist Research Award (No. 2009/20/37/8/BRNS/3328).

**Figure and table caption:**

**Fig. 1** Fluorescence spectra of Acf (1), RhB (2) and Acf + RhB (3) (1:1 volume ratio) in water solution. Striking wavelength was 420 nm (Acf absorption maximum) and concentration of individual dye (pure Acf and RhB) $10^{-6}$ M. Inset (1) shows the normalized absorption spectrum of Rhodamine B and fluorescence spectrum of Acriflavine in water solution and inset (2) shows the excitation spectra for Acf+RhB mixture with emission wavelengths at 500 nm (a) and 578 nm (b).

**Fig. 2** Fluorescence spectra of Acf – RhB (1:1 volume ratio) mixed aqueous solution in absence of DNA (1) and in presence of DNA (2). Striking wavelength was 420 nm (Acf absorption maximum) and concentration of individual dye (pure Acf and RhB) $10^{-6}$ M. DNA concentration was 1 μg/ml.

**Fig. 3** (a) Molecular structure of rhodamine B (b) molecular structure of acriflavine (c) structure of DNA showing the negatively charged phosphate deoxyribose backbone (d) schematic diagram showing the attachment of Acf & RhB onto phosphate backbone of DNA.

**Table 1** Function table of NOT gate using fluorescence intensity

**Table 2** Function table of YES-NOT gate using fluorescence intensity

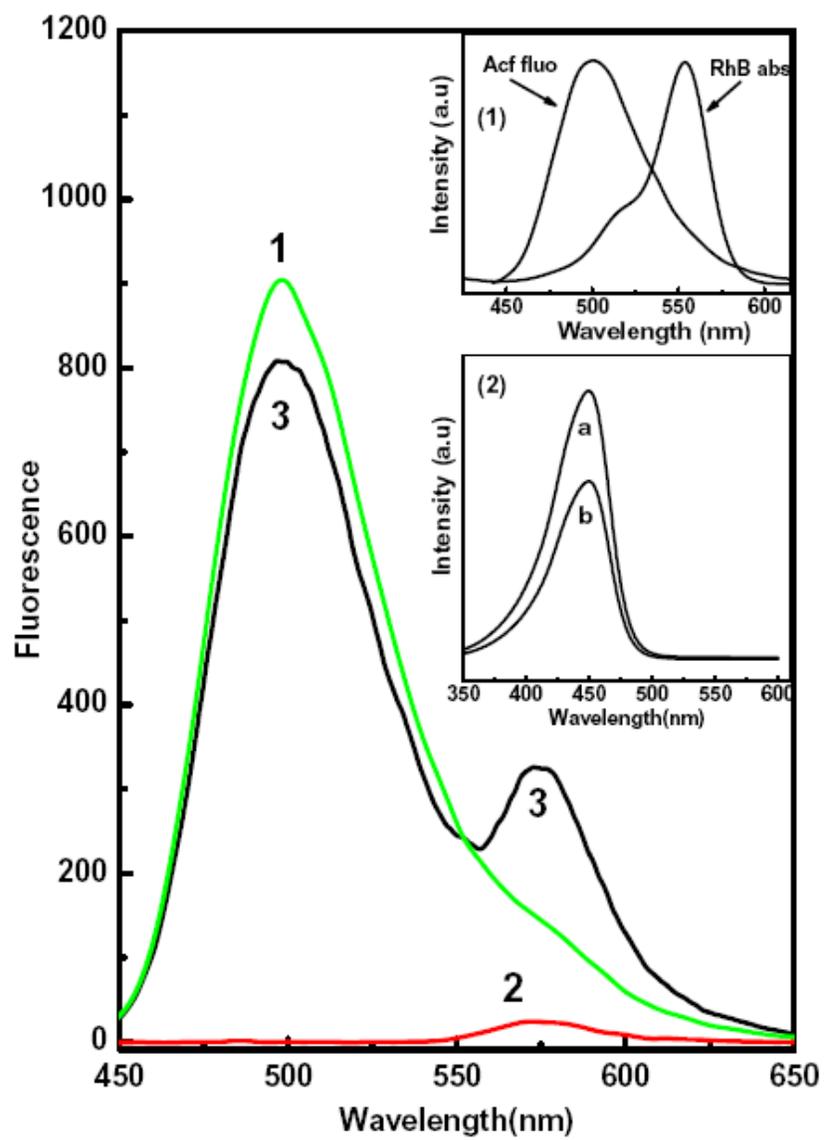

**Fig. 1**

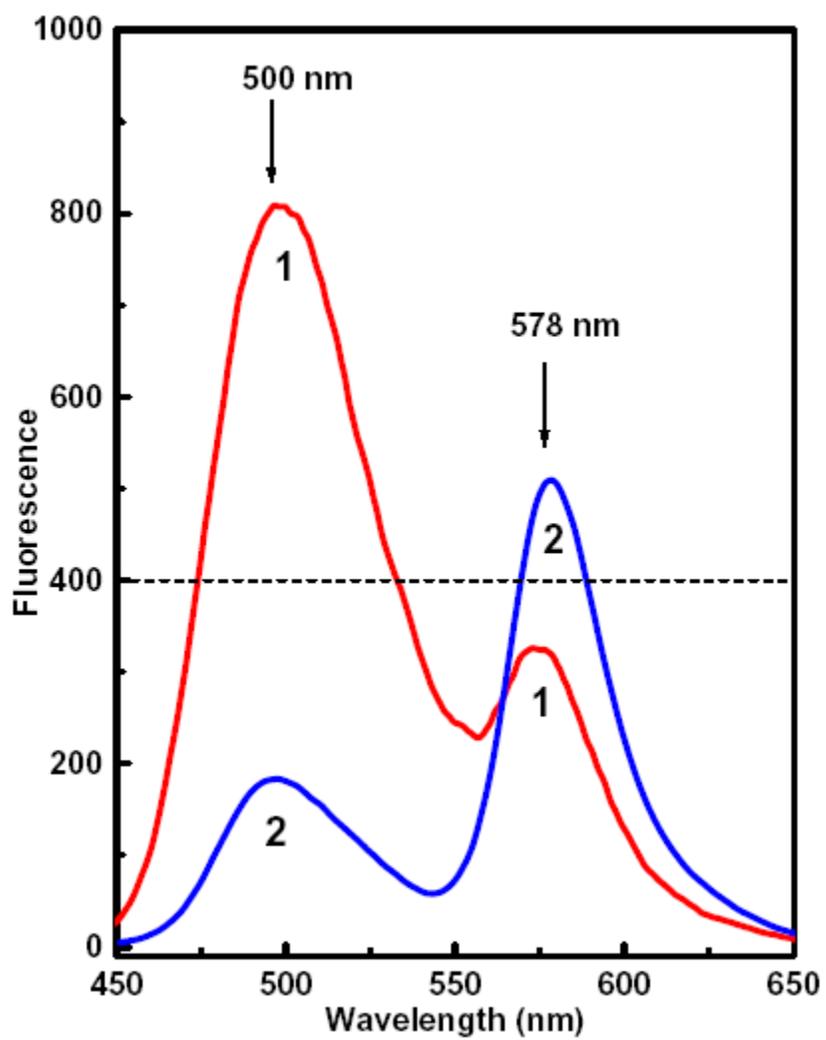

**Fig. 2**

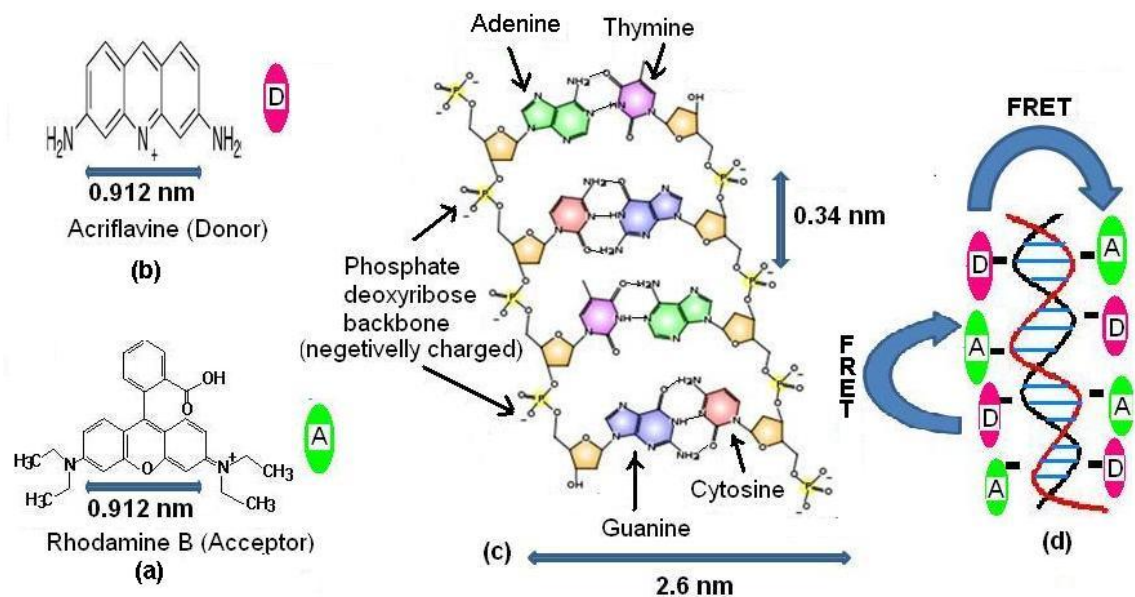

**Fig. 3**

| INPUT DNA | OUTPUT (Fluorescence intensity of 500 nm band) |
|---|---|
| 0 (absence of DNA) | 1 (Fluorescence intensity greater than reference level) |
| 1 (presence of DNA) | 0 (Fluorescence intensity less than reference level) |

**Table 1**

| INPUT DNA | OUTPUT (Fluorescence intensity of 500 nm band) | OUTPUT (Fluorescence intensity of 578 nm band) |
|---|---|---|
| 0 (absence of DNA) | 1 (Fluorescence intensity greater than reference level) | 0 (Fluorescence intensity less than reference level) |
| 1 (presence of DNA) | 0 (Fluorescence intensity less than reference level) | 1 (Fluorescence intensity greater than reference level) |

**Table 2**